\documentclass[aip,jap,reprint]{revtex4-1}
\usepackage{graphicx}
\usepackage{ulem}
\usepackage{amssymb}
\usepackage{amsmath}
\begin{document}


\newcommand{\be}{\begin{equation}}
\newcommand{\ee}{\end{equation}}
\newcommand{\bea}{\begin{eqnarray}}
\newcommand{\eea}{\end{eqnarray}}
\newcommand{\Tbar}{{\bar{T}}}
\newcommand{\En}{{\cal E}}
\newcommand{\K}{{\cal K}}
\newcommand{\U}{{\cal U}}
\newcommand{\GC}{{\cal \tt G}}
\newcommand{\Lop}{{\cal L}}
\newcommand{\DB}[1]{\marginpar{\footnotesize DB: #1}}
\newcommand{\q}{\vec{q}}
\newcommand{\kt}{\tilde{k}}
\newcommand{\Lopn}{\tilde{\Lop}}
\newcommand{\noi}{\noindent}
\newcommand{\ovn}{\bar{n}}
\newcommand{\ovx}{\bar{x}}
\newcommand{\ovE}{\bar{E}}
\newcommand{\ovV}{\bar{V}}
\newcommand{\ovU}{\bar{U}}
\newcommand{\ovJ}{\bar{J}}
\newcommand{\calE}{{\cal E}}
\newcommand{\ovphi}{\bar{\phi}}
\newcommand{\zt}{\tilde{z}}
\newcommand{\nuv}{\rm v}
\newcommand{\ds}{\Delta s}

\title{The tunneling potential for field emission from nanotips}
 

\author{Debabrata Biswas}
\affiliation{
Bhabha Atomic Research Centre,
Mumbai 400 085, INDIA}
\affiliation{Homi Bhabha National Institute, Mumbai 400 094, INDIA}
\author{Rajasree Ramachandran}
\affiliation{
Bhabha Atomic Research Centre,
Mumbai 400 085, INDIA}
\author{Gaurav Singh}
\affiliation{
Bhabha Atomic Research Centre,
Mumbai 400 085, INDIA}
\affiliation{Homi Bhabha National Institute, Mumbai 400 094, INDIA}


\begin{abstract}
  In the quasi-planar approximation of field emission, the potential energy due
  to an external electrostatic field $E_0$
  is expressed as  $-e \gamma E_0 \ds$ where $\ds$ is the
  perpendicular distance from the emission site and $\gamma$ is the local field enhancement factor
  on the surface of the emitter. We show that for curved emitter tips, the current density
  can be accurately computed if terms involving $(\ds/R_2)^2$ and $(\ds/R_2)^3$
  are incorporated in the potential where $R_2$ is the second (smaller) principle radius of
  curvature. The result is established analytically for the hemiellipsoid and hyperboloid  emitters
  and it is found that for sharply curved emitters, the expansion coefficients are equal and coincide with that
  of a sphere. The expansion seems to be applicable to generic emitters as demonstrated numerically for an
  emitter with a conical base and quadratic tip. The correction terms in the potential are adequate for
  $R_a \gtrapprox 2$~nm for local field strengths of $5$~V/nm or higher. The result can also be used for
  nano-tipped emitter arrays or even a randomly placed bunch of sharp emitters.
  \end{abstract}






\maketitle




\section{Introduction}
\label{sec:intro}

Electron beams find applications in a variety of devices that include the microwave as well
as sub-millimeter wave generators and amplifiers, accelerators, microscopes as well for use in lithography, welding,
furnace, medical and space applications \cite{booske2008,whaley2009,booske2011,whaley2014,polk2008,graves2012}.
Common mechanisms for producing an electron beam are
thermionic, field and photo emission. A topic of current research centres around 
large area arrays of pointed field emitters \cite{spindt76,spindt91,forbes2012,zhang2014,harris2015,jap2016} that offer high brightness, high current density beams
having a small spread in energy at low operational temperatures.

Field emission of electrons is commonly studied using a Fowler-Nordheim (FN) type model that
involves a planar metallic surface subjected to
a uniform external electrostatic field $E_0$ and the attendant image force between the electron and its image
due to the grounded metallic plane \cite{FN,Nordheim,murphy,jensen2003,forbes_deane}.
Since electron emission is predicted to be weak in the planar case,
the focus has been on sharp protrusions from such a surface, where field enhancement is known to
occur and can lead to a significant jump in electron emission. An improper surface finish can for example
lead to undesirable dark currents in accelerators while properly grown nanotube arrays on a planar substrate
can be the basis of a high performance cold cathode. In both cases, the protrusions are sharp and only
their tips act as electron emitters. Field emission in such cases is handled by a quasi-planar extension
where the local electric field continues to be uniform across the tunneling region but its magnitude
is enhanced by the field enhancement factor $\gamma$.
However, when the protrusions are sharp and the apex radius of curvature is only a few nanometers, the local
electric field decreases significantly even within the tunneling regime.
This change in local field away from the surface of curved emitters should thus be incorporated as corrections in order to
predict the emitted current density accurately.

The nonlinear nature of the external field near the surface of curved emitters is well known
\cite{cutler93a,cutler93b,fursey,forbes2013}.
For exactly solvable problems such as the hyperboloid, the deviation from the planar result has been
demonstrated in the form of nonlinear FN-plots  and the current densities were found to
differ by orders of magnitude \cite{cutler93a,cutler93b}. In general however,
a first approximation in dealing with curved emitters is to treat
the surface locally as a sphere having the same local radius of curvature. Thus, the external potential
may be expressed locally as \cite{fursey,forbes2013}

\be
V_{ext} \simeq E_l \ds \frac{1}{1 + (\ds/R)}  \label{eq:sph}
\ee

\noi
where $R$ is the local radius of curvature and $\ds$ is the perpendicular distance from the surface.
For  axially symmetric emitters, 
the form of the  nonlinear external potential has recently been studied using a different approach\cite{KX}.
It has been shown that along the symmetry axis of the emitter,
for $\ds < R_a$, the external potential energy $V_{ext}$ takes the form 

\be
V_{ext}^{(a)} (\Delta s)  \simeq E_l \Delta s( 1 - \Delta s/R_a) \label{eq:KX}
\ee

\noi
where $R_a$ is the apex radius of curvature and the normal distance $\Delta s$ is measured from the apex.

The validity of the local spherical approximation can be scrutinized using
exact results for curved emitters such as the hyperboloid or hemi-ellipsoid.
A similar approach for the image potential shows that
for the hyperboloid, where exact results for the image charge potential due to a ring of charges
is known \cite{jensen_image}, the spherical approximation is found to hold \cite{BR2017a}
near the tip of sharp hyperboloids when the local radius of curvature considerably exceeds the tunneling
distance.

The approach that we adopt here makes use of the exact results for the hemi-ellipsoid and hyperboloid
emitters to derive a correction to Eq.~\ref{eq:KX} and determine the conditions under which
it is identical for the two emitters. We then show that an identical result exists for the sphere provided
corrections to Eq.~\ref{eq:sph} are incorporated. Our derivation also brings out the role of the
principle radii of curvature ($R_1,R_2$) and the added clarification that the spherical approximation, where
applicable, must be used with $R_2$ except at the apex where $R_1 = R_2$.
Finally, the applicability of the result is tested numerically for a conical emitter with a quadratic tip and
found to be in good agreement.

\section{Potential variation normal to the surface}

We shall first deal with the potential variation along field lines close to the surface of a
hemiellipsoid and a hyperboloid \cite{kos,pogorelov}. In both cases, the structure is assumed to be
vertically aligned  ($\hat{z}$) in the presence of an external field. It is convenient to work in
{\it prolate spheroidal coordinate} system ($\eta,\xi,\phi$). These are related to the
Cartesian coordinates by the following relations:

\bea
\nonumber
&&x= c_2 \sqrt{({\eta^2}-1)(1-{\xi^2})}\cos{\phi}\\
\nonumber
&&y= c_2 \sqrt{({\eta^2}-1)(1-{\xi^2})}\sin{\phi}\\
&&z= c_2 \xi \eta,
\label{eq:cart_pro_sph}
\eea

\noi
Note that a surface obtained by fixing $\eta=\eta_0$ in this
coordinate system is an ellipsoid while $\xi = \xi_0$ defines a hyperboloid.

For a hemiellipsoid in an external field $-E_0 \hat{z}$, 
the field lines close to the surface are $\xi = constant$ curves.  For a hyperboloid diode
with both the cathode and anode as hyperboloid surfaces, the field lines are always $\eta = constant$ curves.
Further, since we are concerned with potential variation over a distance of around $1~$nm
at moderate fields of $5$~V/nm, we shall assume that the curved field lines are
approximately straight over this distance. Its validity is tested in the appendix
for a hemiellipsoid where it is shown using a Taylor expansion, that close to the apex from where field
emission predominantly occurs, the straightness assumption is largely valid.

\subsection{hemiellipsoid}

Consider a hemiellipsoidal emitter, $\eta = \eta_0$, on a grounded conducting plane,
placed in an external electrostatic field $-E_0 \hat{z}$.
The solution of Laplace equation
may be written as\cite{kos,pogorelov,jap2016}

\begin{equation}
	V(\eta,\xi) = c_2  E_0 \eta \xi \Bigg( 1 - \frac{\log\big[ \frac{\eta+1}{\eta-1} \big] - \frac{2}{\eta}}{\log\big[ \frac{\eta_0+1}{\eta_0-1} \big] - \frac{2}{\eta_0}} \Bigg)	\label{analytic_V}
\end{equation}

\noi
where $c_2 = \sqrt{h(h - R_a)}$, $h$ is the height and $R_a$ is the apex radius of curvature.
The point $(\eta,\xi)$ may lie on the hemiellipsoid surface or outside. We wish to
determine the variation in potential close to the surface along the field line $\xi = \xi_0$ at the point $(\eta_0,\xi_0)$.

Using Eq.~\ref{analytic_V}, the electrostatic potential $V$ at this local
point $(\eta_0+\Delta \eta,\xi_0)$ outside the surface can be calculated as

\be
\begin{aligned}
V( & \eta_0 + \Delta \eta,\xi_0) = {} \U  \bigg[1 - \frac{\log(\frac{\eta_0 + \Delta\eta + 1}{\eta_0 + \Delta\eta - 1})-\frac{2}{\eta_0 + \Delta\eta}}{\log( \frac{\eta_0+1}{\eta_0-1}) - \frac{2}{\eta_0}}\bigg] \nonumber\\
&= - \tilde{\U} \bigg[\frac{2}{\eta_0} + \log\bigg(\frac{1+\frac{\Delta\eta}{\eta_0+1}}{1+\frac{\Delta\eta}{\eta_0 -1}}\bigg)-\frac{2}{\eta_0(1+\frac{\Delta \eta}{\eta_0})}\bigg]			\nonumber\\
& = -\tilde{\U} \bigg[\frac{2}{\eta_0}\Big[\Big(\frac{\Delta \eta}{\eta_0}\Big) - \Big(\frac{\Delta \eta}{\eta_0}\Big)^2 + \Big(\frac{\Delta \eta}{\eta_0}\Big)^3 +  \ldots\Big] \nonumber\\
    &~~~~~ + \log\Big(1+\frac{\Delta\eta}{\eta_0+1}\Big) - \log{\Big(1+\frac{\Delta\eta}{\eta_0 -1}}\Big)\bigg]
\end{aligned}
  \ee

\noi
where

\bea
\U & = & c_2  E_0 \xi_0 \eta_0 + c_2  E_0 \xi_0 \Delta\eta = \U_0 + \Delta\U \\
\tilde{\U} &  = & \frac{\U_0 + \Delta\U}{\log( \frac{\eta_0+1}{\eta_0-1}) - \frac{2}{\eta_0}} =
\tilde{\U_0} + \tilde{\Delta\U}.
\eea

\noi
Using the expansion $\log(1+x) = x - x^2/2 + x^3/3 + \ldots$ the above expression
for the potential can be approximated  as

\be
\begin{aligned}
  V(&\eta_0 + \Delta \eta,\xi_0) \simeq
  2\tilde{\U} \bigg[ \frac{1}{\eta_0^2(\eta_0^2 - 1)}\Delta\eta ~+  \\
    &\Big(\frac{1}{\eta_0^3}-\frac{\eta_0}{(\eta_0^2 -1)^2}\Big)(\Delta\eta)^2
 - \Big( \frac{1}{\eta_0^4} - \frac{3\eta_0^2 + 1}{3(\eta_0^2-1)^3}\Big)(\Delta \eta)^3 \bigg] \nonumber
\end{aligned}
\ee

\noi
which on simplifying and keeping terms upto $(\Delta\eta)^3$, takes the form

\be
\begin{aligned}
  V(\eta_0+\Delta \eta,\xi_0) \simeq {} & \frac{2\tilde{\U_0}\Delta\eta}{(\eta_0^2-1)\eta_0^2}~ \times \\
  & \bigg[1-\frac{\eta_0 }{\eta_0^2-1}\Delta\eta + \frac{4\eta_0^2}{3(\eta_0^2-1)^2} (\Delta \eta)^2\bigg]    \nonumber \label{eq:Vdeta} \\
\end{aligned}
\ee

\noi
Rewriting in terms of magnitude of the local field $E_{l}$ 
\be
\begin{aligned}
E_{l}(\eta_0,\xi_0) =\frac{2E_0 \xi_0}{\eta_0 \sqrt{\eta_0^2 - \xi_0^2}\sqrt{\eta_0^2 - 1} (\log\big[ \frac{\eta_0+1}{\eta_0-1} \big] - \frac{2}{\eta_0})}
\end{aligned}
\ee

\noi
and the normal distance $\Delta s$ from the point $(\eta_0,\xi_0)$

\be
\Delta \eta = \frac{\Delta s}{h_\eta} + \mathcal{O} ((\Delta s)^2) \label{eq:s}
\ee

\noi
where

\be
h_\eta =  c_2\sqrt{\frac{\eta_0^2-\xi_0^2}{\eta_0^2-1}}, \label{eq:heta}
\ee

\noi
the potential $V(\eta_0+\Delta \eta,\xi_0)$  can
be expressed as

\be
\begin{aligned}
 V(\Delta s) & \simeq {}  E_{l}(\eta_0,\xi_0) \Delta s
  \bigg[1 - \frac{\Delta s}{R_1} \frac{(\eta_0^2-\xi_0^2)}{(\eta_0^2-1)}~+ \\
  & \frac{4}{3}\Big(\frac{\Delta s}{R_1}\Big)^2 \Big(\frac{\eta_0^2-\xi_0^2}{\eta_0^2-1}\Big)^2 \bigg]. \label{eq:ellip_pre}
\end{aligned}
\ee

\noi
For an ellipsoid $\eta = \eta_0$, the principal local radii of curvature at the point ($\eta_0,\xi_0$)
are

\bea
R_1 & = & R_a \frac{\big(\eta_0^2 - \xi_0^2\big)^{3/2}}{\big(\eta_0^2 - 1\big)^{3/2}} \\
R_2 & = &  R_a \frac{\big(\eta_0^2 - \xi_0^2\big)^{1/2}}{\big(\eta_0^2 - 1\big)^{1/2}}
\eea

\noi
while the Gaussian radius of curvature is

\be
R_g = (R_1 R_2)^{1/2} = R_a \frac{\eta_0^2 - \xi_0^2}{\eta_0^2 - 1}.
\ee

\noi
Thus, Eq.~\ref{eq:ellip_pre} can be  further simplified as

\be
\begin{aligned}
	V(\Delta s) \simeq E_{l}(\eta_0,\xi_0) \Delta s\Bigg
	[ 1-\bigg(\frac{\Delta s}{R_2}\bigg) + \frac{4}{3}\bigg(\frac{\Delta s}{R_2}\bigg)^2\Bigg]
        \label{eq:ellip_final}
\end{aligned}
\ee

\noi
and forms the central result of this paper. It can be used to estimate the tunneling transmission coefficient and
hence the current density at a point close to the emitter apex.

Note that Eq.~\ref{eq:ellip_final} represents approximately the potential variation along the normal to a point
on the surface of the hemiellipsoid. However, in the apex neighbourhood of a sharp emitter, Eq.~\ref{eq:ellip_final} does represent
the normal potential variation close to the surface quite accurately (see appendix) and can thus be used to determine
emission currents.

\subsection{Hyperboloid}

The hyperboloid emitter surface is defined by $\xi = \xi_0 = \sqrt{D/(D + R_a)}$ while a flat anode $\xi = 0$ is placed
a distance $D$ below the tip. In the transformation equations of Eq.~\ref{eq:cart_pro_sph},
$c_2 = \sqrt{D(D + R_a)}$ where $R_a$ is the apex radius of curvature \cite{hyperbola}.
The derivation of the potential variation follows a similar line. If the potential difference between
the anode and cathode is $V_0$, the potential at any point can be expressed as

\be
V(\eta,\xi) = V_0~\Bigg( 1 - \frac{\ln\Big[\frac{1~ -~ \xi}{1~ +~ \xi}\Big]}{\ln\Big[\frac{1~ -~ \xi_0}{1~ + ~\xi_0}\Big]} \Bigg)
\ee

\noi
Thus for small excursions along the field line $\eta = \eta_0$ starting from the point ($\eta_0,\xi_0$) on the hyperboloid
surface, the potential

\be
\begin{aligned}
V(\eta_0,\xi_0 - \Delta\xi)  = {} -\frac{V_0}{\ln\big(\frac{1 - \xi_0}{1 + \xi_0}\big)}\Big[ & \ln\big(1 +
  \frac{\Delta\xi}{1 - \xi_0}\big) - \\
  & \ln\big(1 - \frac{\Delta\xi}{1 + \xi_0}\big) \Big]
\end{aligned}
\ee

\noi
Keeping terms upto $(\Delta\xi)^3$, we have

\be
\begin{aligned}
  V(\eta_0, \xi_0 - \Delta\xi)   \simeq  & - \frac{2V_0}{\ln\big(\frac{1 - \xi_0}{1 + \xi_0}\big)} \frac{\Delta\xi}{1 - \xi_0^2} \Big[
    1 - \frac{\xi_0}{1 - \xi_0^2}\Delta\xi \\
  &  + \frac{1 + 3\xi_0^2}{3(1 - \xi_0^2)^2} (\Delta\xi)^2  \Big].
\end{aligned}
\ee

\noi
In terms of the normal distance

\be
\Delta s = c_2 \Delta\xi \sqrt{\frac{\eta_0^2 - \xi_0^2}{1 - \xi_0^2}}
\ee

\noi
and the local electric field

\be
E_{l} = - \frac{V_0}{c_2} \frac{1}{(1 - \xi_0^2)} \frac{2}{\ln\big[\frac{1 - \xi_0}{1 + \xi_0}\big]}
\ee

\noi
the potential variation $V(\eta_0, \xi_0 - \Delta\xi)$ can be expressed as a function of $\Delta s = h_\xi \Delta \xi$ as

\be
V(\Delta s) \simeq  E_{l} \Delta s \Big[ 1 - \frac{\Delta s}{R_2} + \frac{1 + 3\xi_0^2}{3 \xi_0^2} \bigg(\frac{\Delta s}{R_2}\bigg)^2 \Big]  
\ee

\noi
where $R_2 =  R_aR_{1}/R_g$ is a principal radius of curvature for the hyperboloid $\xi = \xi_0$ evaluated at the point ($\eta_0,\xi_0$).
The respective radii of curvature can be expressed as

\bea
R_1 & = & R_a \frac{\big(\eta_0^2 - \xi_0^2\big)^{3/2}}{\big(1 - \xi_0^2\big)^{3/2}} \\
R_2 & = &  R_a \frac{\big(\eta_0^2 - \xi_0^2\big)^{1/2}}{\big(1 - \xi_0^2\big)^{1/2}} \\
R_g & = & (R_1 R_2)^{1/2} = R_a \frac{\eta_0^2 - \xi_0^2}{1 - \xi_0^2}.
\eea

\noi
For a reasonably sharp emitter tip, $\xi_0$ is
close to unity. As an illustration, for $D = 5000$nm and $R_a = 5$nm, $\xi_0 = 0.99950$ while for
$D = 1500$nm and $R_a = 5$nm, $\xi_0 = 0.99834$. Thus, setting $\xi_0$ to be 1,

\be
V(\Delta s) \simeq  E_{l} \Delta s \Big[ 1 - \frac{\Delta s}{R_2} + \frac{4}{3} \bigg(\frac{\Delta s}{R_2}\bigg)^2 \Big] \label{eq:finalpot}
\ee

\noi
as in the case of hemiellipsoid. As before, Eq.~\ref{eq:finalpot} is more accurately the
potential variation along field lines of constant $\eta = \eta_0$ and only approximately so along
the normal distance. Close to the tip of a sharp hyperboloid however, it is expected that Eq.~\ref{eq:finalpot} is a good
approximation for the potential variation normal to the emitter near the apex.

\subsection{The Sphere}

For a grounded conducting sphere of radius $R$ in an electric field $-E_0 \hat{z}$, the potential
outside the sphere is

\be
V_{ext} = E_0 r \cos\theta  \big[1 - \frac{R^3}{r^3}\big].
\ee

\noi
Writing $r = R + \ds$,

\be
V_{ext} = E_0 \cos\theta \Big[ \frac{3R\ds}{R + \ds} + \frac{(\ds)^3}{(R + \ds)^2} \Big].
\ee

\noi
Writing $E_l = 3E_0 \cos\theta$ and neglecting the second term leads us to Eq.~\ref{eq:sph}.
However, since we are interested in a correction term of the order of $(\ds)^3$, the second
term must be retained. Now assuming $\ds/R << 1$,

\bea
V_{ext} & \simeq & E_l \ds \Big [\big\{1 - \frac{\ds}{R} + \big(\frac{\ds}{R}\big)^2 \big\} + \frac{1}{3}\big(\frac{\ds}{R}\big)^2 \big] \\
& = & E_l \ds \Big [1 - \frac{\ds}{R} + \frac{4}{3} \big(\frac{\ds}{R}\big)^2 \Big]
\eea

\noi
which is identical to the result obtained above for the hemi-ellipsoid and the hyperboloid.

\subsection{Generic emitter tips}

A derivation of a corrected formula for the external potential variation
applicable to generic emitter is not
readily available. However, we shall investigate the applicability of Eq.~\ref{eq:finalpot}
for generic emitters with parabolic tips. Note that cylindrically symmetric emitter tips
that are vertically aligned can be approximated as
\bea
z &  = & h  + \frac{1}{2} \Big(\frac{d^2 z}{d\rho^2}\Big)_{\rho = 0} \rho^2  + \ldots \\
& \simeq &  h\Big[1 - \frac{1}{2} \frac{\rho}{R_a}\frac{\rho}{h} \Big] \label{eq:quadratic}
\eea

\noi
where $R_a$ is the magnitude of the apex radius of curvature, $\rho = (x^2 + y^2)^{1/2}$, $h$ is the height
of the emitter and we have assumed that the tip is not flat ($(d^2 z/d\rho^2)_{\rho = 0} \neq 0$).
Also, since field emission occurs close to the tip, higher order terms in $\rho$ can
be ignored in the expansion of $z$.

Eq.~\ref{eq:quadratic} can  be used to find the local and gaussian curvatures in terms of the
apex radius of curvature. Moreover, recent results \cite{B2017b} show that local surface electric field 
around the tip can be expressed in terms of the local electric field at the apex ($E_a$) and a
generalized $\cos\tilde{\theta}$ factor:

\be
E_{l}(z) = E_a \cos\tilde{\theta} = E_0 \frac{\gamma_a (z/h)}{\sqrt{(z/h)^2 + (\rho/R_a)^2}} \label{eq:localE}
\ee

\noi
where $\gamma_a$ is the field enhancement factor at the apex and $z$ is the height on the emitter surface
measured from the conducting plane.

For a surface parameterized as ($\rho\cos\varphi,\rho\sin\varphi,h - a\rho^2$), where $a = 1/(2R_a)$
and $\rho = (x^2 + y^2)^{1/2}$, the
local Principal and Gaussian radii of curvature are respectively

\bea
R_1 & = &  -R_a \bigg[1 + \big(\frac{\rho}{R_a}\big)^2\bigg]^{3/2} \label{eq:r1} \\
R_2 & = &  -R_a \bigg[1 + \big(\frac{\rho}{R_a}\big)^2\bigg]^{1/2} \label{eq:r2} \\
R_g & = &   R_a \bigg[1 + \big(\frac{\rho}{R_a}\big)^2\bigg] \label{eq:rg}
\eea

\noi
Thus, for quadratic emitters, $E_{l}$ and $R_2$ in Eq.~\ref{eq:finalpot} 
are given by Eq.~\ref{eq:localE} and \ref{eq:r2} respectively if the apex radius of curvature and
field enhancement factors are known. Alternately, they can be computed at each point on the
emitter surface if the exact numerical solution
for the potential is available.
As in case of the hyperboloid, Eq.~\ref{eq:finalpot} is expected to hold for general quadratic emitters
that are sharp.

\section{Numerical Results}

We shall first make a crude estimate of the domain of validity of Eq.~\ref{eq:finalpot}
for a typical local electric field $E_{l} \simeq 5 \times 10^9$ V/m. For an emitter with
work function of $4.5$~eV, the tunneling distance at this
local field is about $1$nm. At the apex, with $R_a = 5$nm, the quadratic term is about $20\%$ of the linear
while the cubic is about $5\%$ of the linear. Thus, along the symmetry axis, the neglect of terms higher than
cubic appears justified when $R_a > 5$nm. 

Away from the emitter apex, the principle radius of curvature $R_2$ increases (albeit slowly compared to $R_1$) for
typical quadratic tips. At the same time, the local electric field decreases for a given external
electric field. Thus, while the tunneling distance increases marginally, the domain of validity of Eq.~\ref{eq:finalpot}
also increases. In the following, we shall explore the difference between the exact current density
and the one obtained using the approximate potential of Eq.~\ref{eq:finalpot}, for various emitter
shapes and position.

\begin{figure}[htb]
\vskip -2.1 cm
\hspace*{-1.0cm}\includegraphics[width=0.6\textwidth]{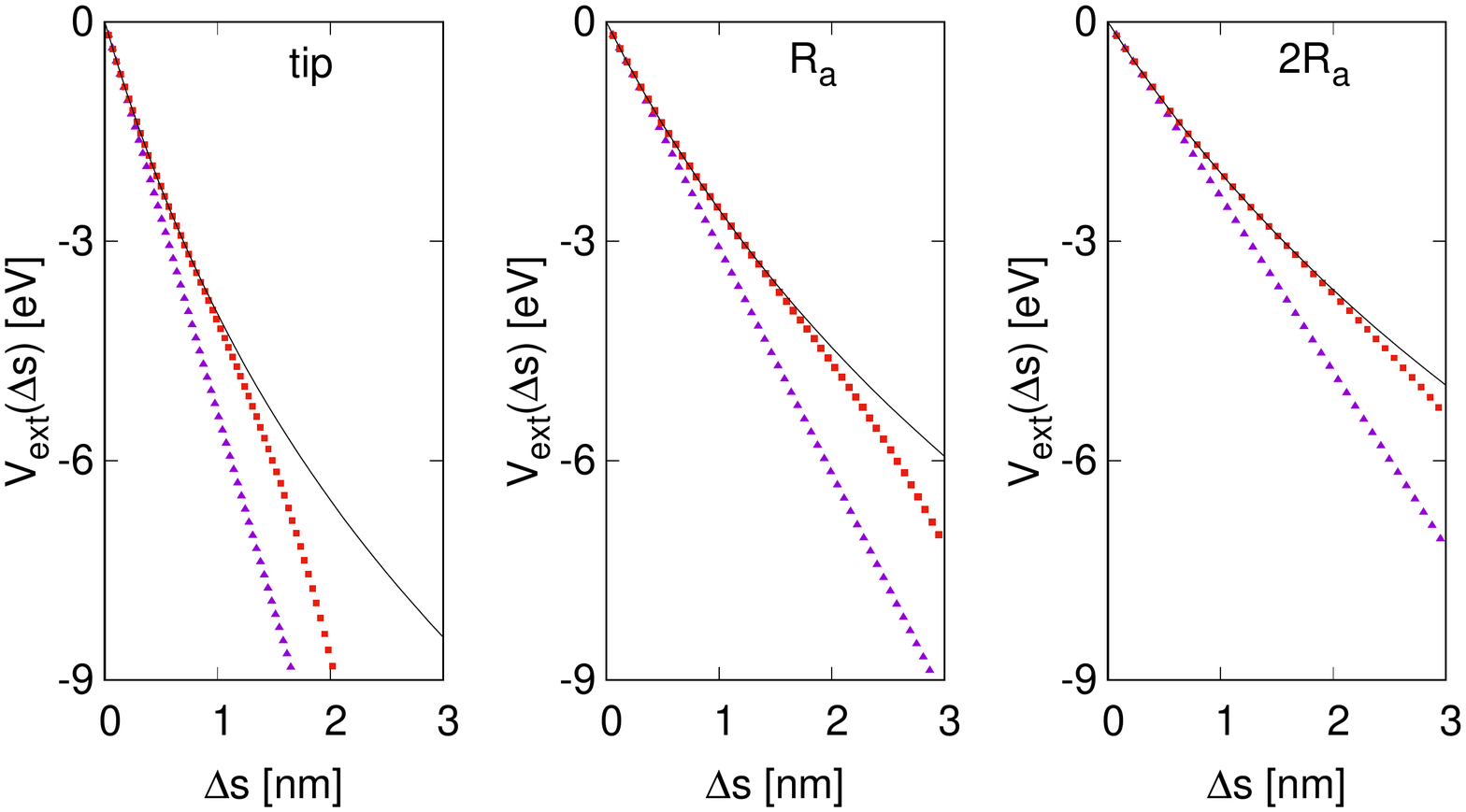}
\vskip -0.6 cm
\caption{The potential energy due to the external field along the normal to three different points on a hemiellipsoidal emitter surface located
  (i) at the tip ($z = h$) (ii) at $z = h - R_a$  (iii) at $z = h - 2R_a$. The external field strength is
  $E_0 = 6\times 10^4$~V/m. The height of the hemiellipsoid $h = 1500~\mu$m while the base radius $b = 2~\mu$m.
  The filled triangles are the quasi-planar result ($-E_{l} \Delta s$), the filled squares are obtained
  using Eq.~\ref{eq:finalpot} while the solid curve is the exact result.
}
\label{fig:pot_ellip_b2}
\end{figure}

\begin{figure}[htb]
\vskip -2.1 cm
\hspace*{-1.0cm}\includegraphics[width=0.6\textwidth]{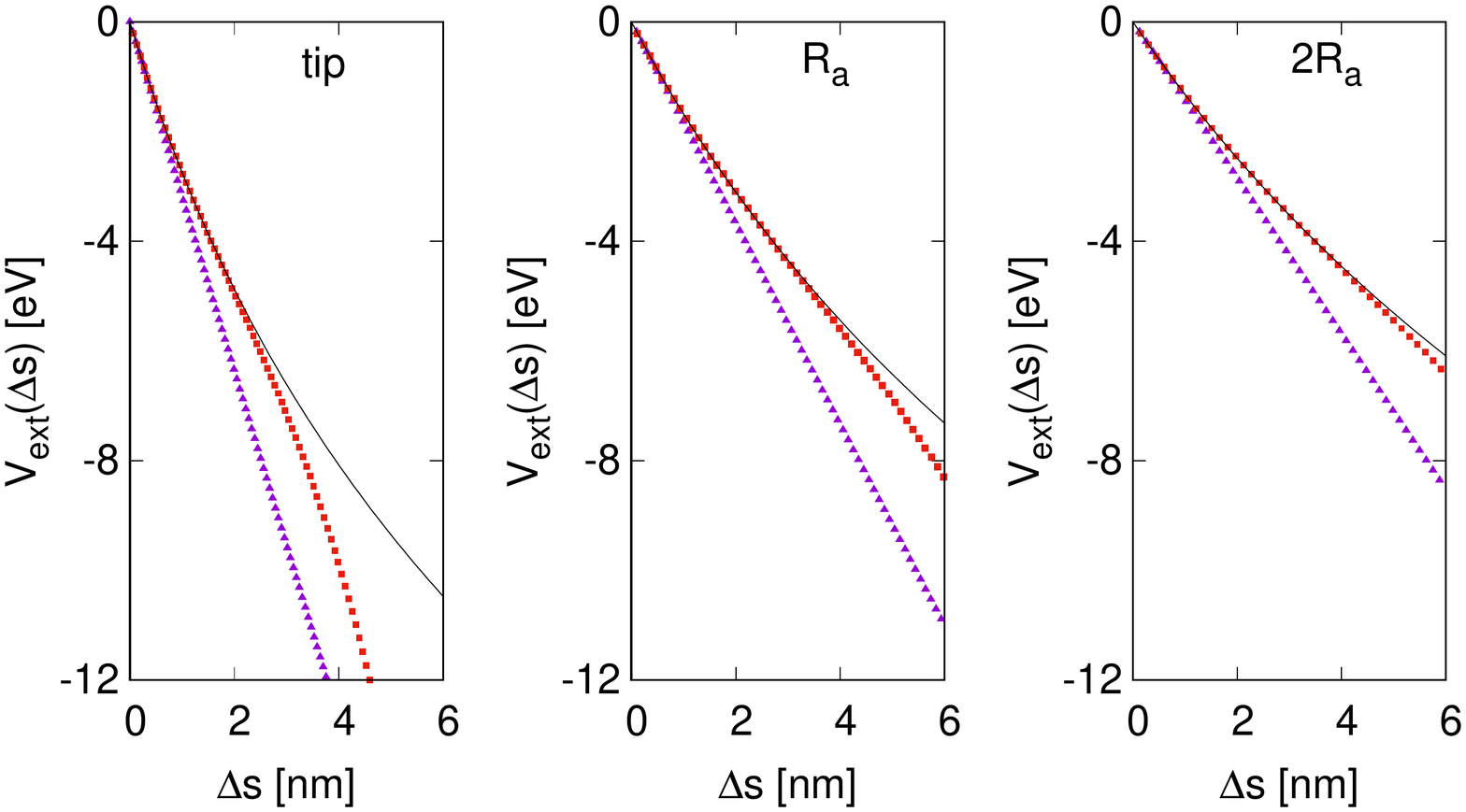}
\vskip -0.6 cm
\caption{As in Fig.~\ref{fig:pot_ellip_b2} for $R_a = 6$nm, 
  $E_0 = 7.5\times 10^4$~V/m and base radius $b = 3~\mu$m.
}
\label{fig:pot_ellip_b3}
\end{figure}

First, we consider a hemiellipsoidal emitter on a grounded conducting plane placed in a uniform electric field.
Fig.~\ref{fig:pot_ellip_b2} shows the potential energy due to the external field at three locations on the emitter surface
(i) at the tip (ii)  at $z = h - R_a$  (iii) at $z = h - 2R_a$.
At the tip  where $R_2  = 2.67$nm, the exact potential and Eq.~\ref{eq:finalpot} match quite well to about $1$nm
while at locations (ii) and (iii) the agreement gets better since $R_2$ increases.
Fig.~\ref{fig:pot_ellip_b3} shows a similar plot for
$R_a = 6$nm and $E_0 = 7.5 \times 10^4$~V/m. The agreement at all three location now gets better.
For an even larger apex radius $R_a = 16.67$nm,
the agreement extends beyond $4$nm at all three locations.

We next turn our attention to the tunneling current densities generated using these potentials. 
Assuming a free electron model, the current density is evaluated at zero temperature as

  \be
  J = \frac{2me}{(2\pi)^2 \hbar^3} \int_0^{E_F} T({\cal E})(E_F  - {\cal E}) d{\cal E}
  \ee

  \noi
  where $T({\cal E})$ is the transmission coefficient at energy ${\cal E}$,
  $m$ is the mass of the electron, $e$ is the
  magnitude of the electron charge and $E_F$ is the Fermi level.  Instead of using the WKB expression for
  the transmission coefficient, we shall determine $T({\cal E})$ numerically 
  using suitable boundary conditions for the 1-dimensional Schr\"{o}dinger equation and
  a modified transfer matrix method \cite{DBVishal}. In the results presented here, curvature corrections to the image potential
  have been neglected in order to bring out the role of corrections to the external potential.

\begin{figure}[htb]
\vskip -2.1 cm
\hspace*{-1.0cm}\includegraphics[width=0.6\textwidth]{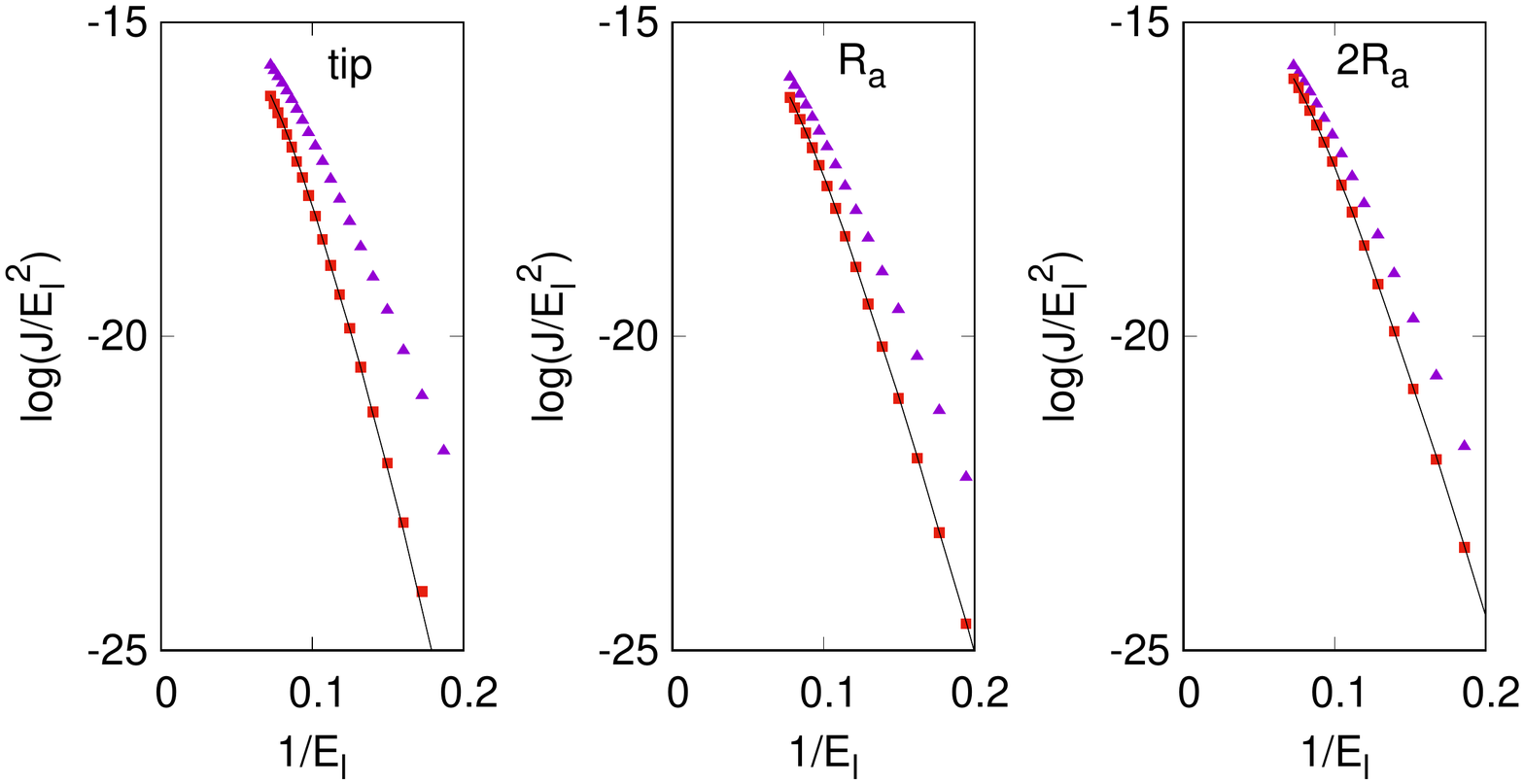}
\vskip -0.6 cm
\caption{A Fowler-Nordheim plot of the current density for a hemiellipsoid with base radius $b = 2\mu$m
  at the three different locations mentioned in Fig.~\ref{fig:pot_ellip_b2}. The solid line is the exact
  result while the filled-squares are obtained using Eq.~\ref{eq:finalpot} for the external potential.
  The filled-triangles are obtained using the quasi-planar approximation for the external potential.
  Here, ${\rm 1/E}_l$ is expressed in the unit $[{\rm V/nm}]^{-1}$.
}
\label{fig:J_ellip_b2}
\end{figure}

\begin{figure}[tb]
\vskip -2.1 cm
\hspace*{-1.0cm}\includegraphics[width=0.6\textwidth]{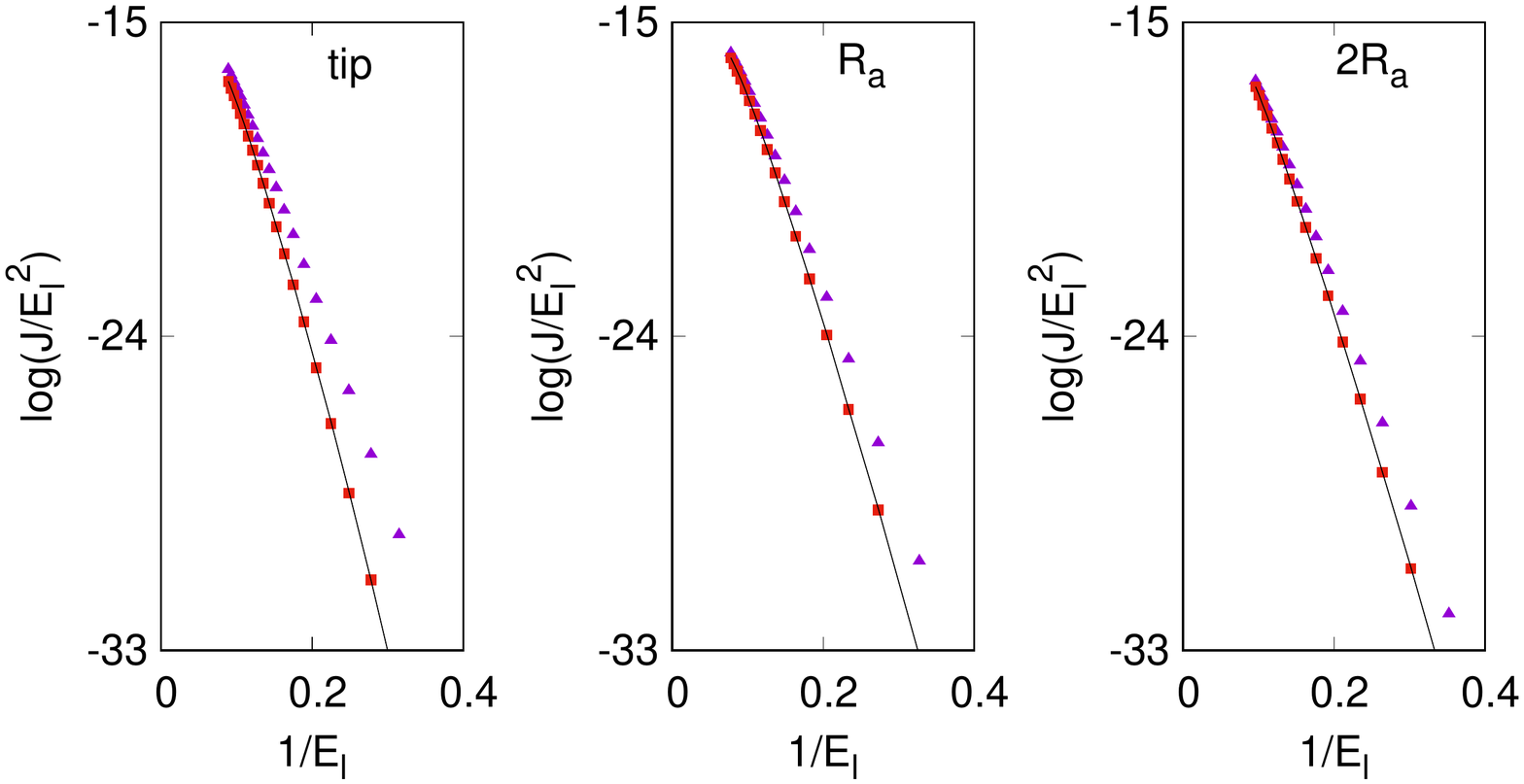}
\vskip -0.6 cm
\caption{As in Fig.~\ref{fig:J_ellip_b2} for $b = 3\mu$m ($R_a = 6$nm)
  at the three different locations. The solid line is the exact
  result while the filled-squares are obtained using Eq.~\ref{eq:finalpot} for the external potential.
  The filled-triangles are obtained using the quasi-planar approximation for the external potential.
}
\label{fig:J_ellip_b3}
\end{figure}

The corresponding current densities for $R_a = 2.67$nm are shown in Fig.~\ref{fig:J_ellip_b2} at the locations
mentioned earlier. Clearly the two correction terms in the potential (see Eq.~\ref{eq:finalpot}) are adequate 
to reproduce the exact results. For $b = 3\mu$m ($R_a = 6$~nm), the current density is shown in
Fig.~\ref{fig:J_ellip_b3}. The agreement
with the exact result remains excellent using Eq.~\ref{eq:finalpot} while the
agreement between the exact and quasi-planar case improves considerably as expected.

We next turn our attention to a case where the analytical solution for the potential is not known.
Using a suitable nonlinear line-charge of height $L$ placed on a grounded conducting plane in the
presence of a uniform electric field, a conical zero-potential surface is obtained of height $300~\mu$m,
base radius $16~\mu$m, having a rounded top with an apex radius of curvature $R_a = 4.56$nm.
The emitter tip is modeled very well\cite{B2017b} by the quadratic $z = h - \rho^2/(2R_a)$.

\begin{figure}[htb]
\vskip -2.1 cm
\hspace*{-1.0cm}\includegraphics[width=0.6\textwidth]{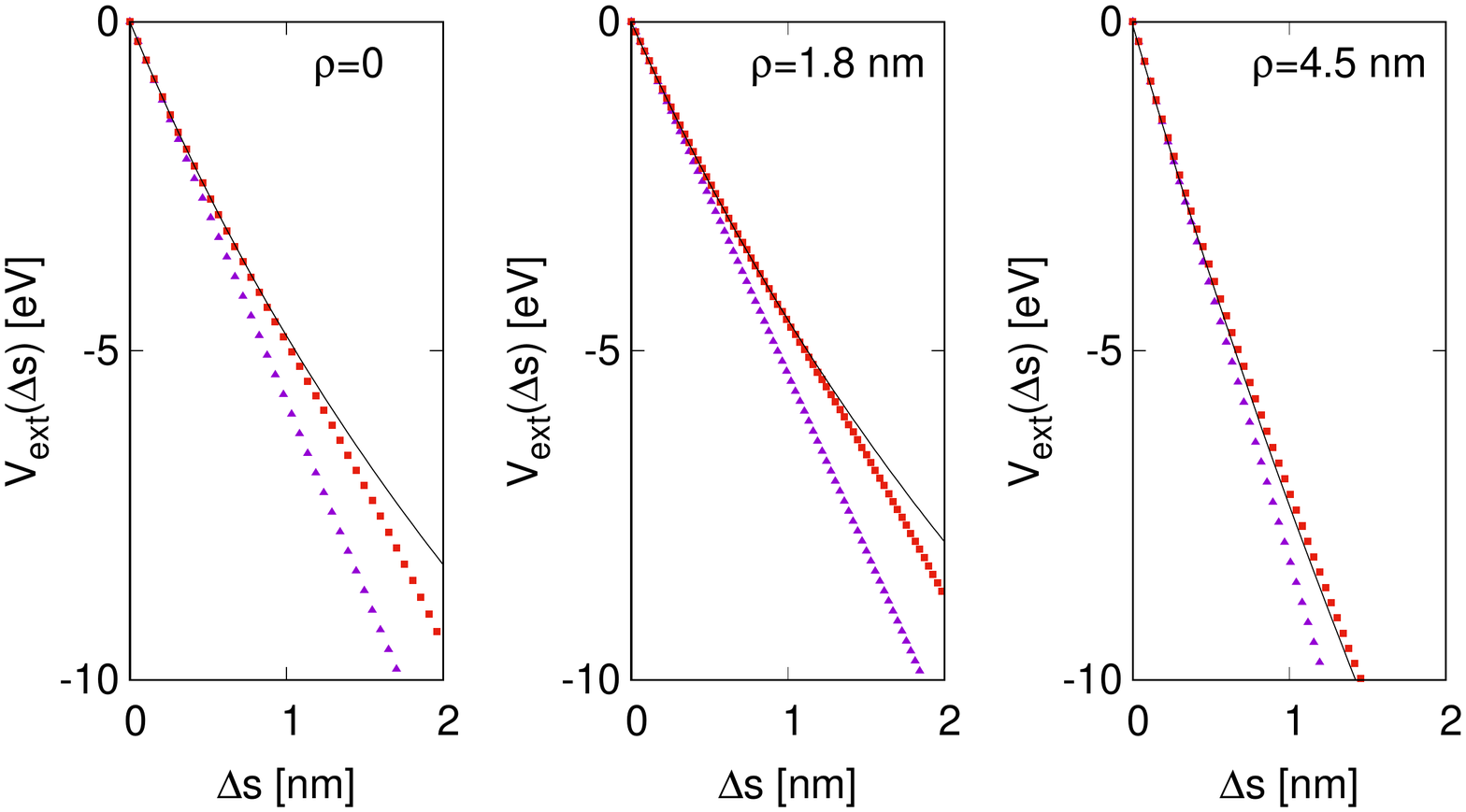}
\vskip -0.6 cm
\caption{The potential energy due to the external field along the normal to points on a rounded conical
  surface located (i) at the tip ($\rho \simeq 0$nm) (ii) at $\rho \simeq 1.8$nm and
(iii) at $\rho \simeq 4.5$~nm.  The external field 
  $E_0 = 5\times 10^5$~V/m. The values of the field enhancement factor at these points are 
  11555, 10730 and 8160 respectively. The filled triangles are the quasi-planar result (infinite
  radius of curvature) while the filled squares are obtained using Eq.~\ref{eq:finalpot}.
  The solid curve is the exact result.
}
\label{fig:pot_cone}
\end{figure}

Fig.~\ref{fig:pot_cone} is a plot of the potential energy variation along the normal to the
emitter surface. The points (from left to right) are located at (i) $\rho = 0$ (the emitter tip)
(ii) $\rho \simeq 1.8$ nm and (iii) $\rho \simeq 4.5$~nm. The exact potential is calculated using the
line charge distribution. Clearly, Eq.~\ref{eq:finalpot} provides a fair approximation to the
exact potential in the tunneling regime.  It gets marginally better away from the apex due to the
increase in $R_2$ but thereafter minor deviations in the tunneling region occur, perhaps
due to the uncertainty in the 4/3 multiplying factor.

\begin{figure}[htb]
\vskip -2.1 cm
\hspace*{-1.0cm}\includegraphics[width=0.6\textwidth]{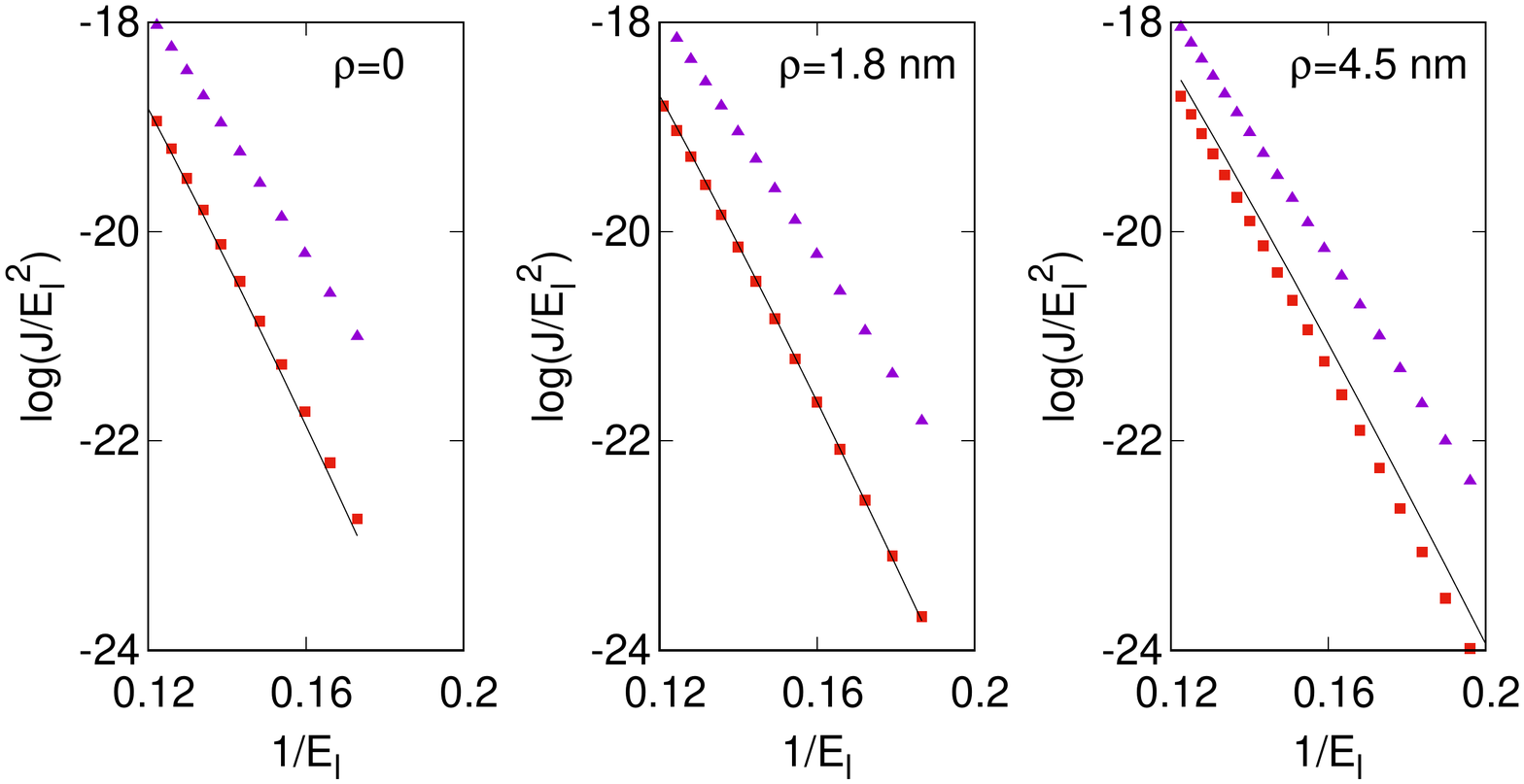}
\vskip -0.6 cm
\caption{The current density as a function of the local electric field $E_l$ at three points
  on the rounded conical tip as mentioned in Fig.~\ref{fig:pot_cone}.
  Here, ${\rm 1/E}_l$ is expressed in the unit $[{\rm V/nm}]^{-1}$.
}
\label{fig:J_cone}
\end{figure}

The corresponding current densities are shown as a Fowler-Nordheim plot in Fig~\ref{fig:J_cone}.
In the first two cases, the current densities using Eq.~\ref{eq:finalpot} for the external
potential are in good agreement with the
exact result (solid line) obtained using the nonlinear line charge distribution. In the third
case (plot to the right), Eq.~\ref{eq:finalpot} underestimates the current density marginally.
The difference with the quasi-planar case is again substantial especially at
smaller values of local field $E_l$, in all three cases. 

In order to determine the effectiveness of Eq.~\ref{eq:finalpot} in determining the total
electron current from a single emitter, we have computed the emitter current at two values of
the external field, $E_{0}$. At $E_0 = 5 \times 10^5$~V/m
(corresponding to a local apex field $\simeq 5.77 \times 10^9$~V/m), the currents obtained from the
quasi-planar approximation, Eq.~\ref{eq:finalpot} and the exact potential are $0.302~\mu$A, $0.0455~\mu$A and
$0.0445~\mu$A respectively. While the last two values are close, the quasi-planar current is nearly 7 times
more. At the higher external field $E_0 = 10^6$~V/m, the currents obtained from the
quasi-planar approximation, Eq.~\ref{eq:finalpot} and the exact potential are 0.36~mA, 0.205~mA and
0.220~mA respectively. The last two values are still close while the quasi-planar approximation
improves considerably.

\section{Discussion and Conclusions}

The study of two analytically solvable models, the hemiellipsoid on a conducting plane and
the hyperboloid diode, led us to Eq.~\ref{eq:finalpot}. In both cases, when the emitter is sharp,
Eq.~\ref{eq:finalpot} is accurate near the apex for short excursions in the normal direction.
An identical result was derived for a sphere thereby establishing that the spherical
approximation for curved emitters (where applicable) must be used with the the second
principle radius of  curvature $R_2$ as the radius of the sphere.

Finally, we have also numerically explored the validity of Eq.~\ref{eq:finalpot} for
an analytically unsolvable case, the cone with a quadratic tip.
Our numerical studies show that in all the examples, the current densities obtained using Eq.~\ref{eq:finalpot}
agree well with the exact result near the emitter tip and show a considerable improvement
compared to the quasi-planar case in predicting the emitter current. At low external field strengths,
where the difference with the quasi-planar case is almost an order of magnitude,
Eq.~\ref{eq:finalpot} predicts the current with less than $2\%$ error. At higher field
strengths, the quasi-planar result improves but is still poor compared to the prediction
of Eq.~\ref{eq:finalpot}.
The results presented here have also been tested for a cylindrical emitter with a quadratic
tip.

While the preceding discussion has centred around a single sharp emitter, it is clear that the
form of the external potential remains the same even if the emitter is part of a regular array or
a randomly distributed bunch of emitters, so long as the emitter tip is smooth and parabolic.
For a bunch of emitters with identical height and apex radius of curvature, the only quantity in Eq.~\ref{eq:finalpot}
that depends on the neighbourhood is the local external electric field, $E_{l}$. This is determined
by the extent of shielding which must be determined separately before calculating the field emission current.

In conclusion, the quasi-planar approximation to the potential due to the external field leads to
large errors in emitted current when the apex radius of curvature $R_a \lessapprox 20$~nm
and the applied external field is small. For $ R_a \gtrapprox 2~\rm{nm} $, Eq.~\ref{eq:finalpot} seems to
provide a very good approximation to the external potential and accurately reproduces the emitter current.
Finally, in addition to curvature effects in the external potential, corrections to the image
potential are also important and must be included in determining the emitter current.

\section{Acknowledgement}

The authors acknowledge several useful discussions with Dr. Raghwendra Kumar.

\section{Appendix}

For orthogonal co-ordinate
systems in which the Laplace and Schr\"odinger equations are separable, tunneling transmission coefficients along field lines can be calculated
using the standard 1-d formalisms. In the general case however, curved field lines would necessitate use of the multi-dimensional tunneling
formalism. In view of a possible general applicability of Eq.~\ref{eq:ellip_final} to non-separable systems, we have
instead chosen to express the external potential in terms of the normal distance $\Delta s$ so that standard 1-dimensional
tunneling results can be used. This also leaves open the possibility of incorporating the results of this paper in a modified
Fowler-Nordheim equation. 

The assumption so far in using the normal distance $\Delta s$ has been that field lines are more or less straight over the tunneling
distance of about $1$~nm at moderate local field strengths. In the following, we shall test this assumption for a hemiellipsoid by Taylor
expanding the potential along the normal direction and comparing with Eq.~\ref{eq:ellip_final}.

Consider a point ($\eta_0,\xi_0$) on the hemiellipsoid $\eta = \eta_0$. A point outside, at a distance $\Delta s$  normal
to the hemiellipsoid at ($\eta_0,\xi_0$) can be written as

\bea
z_1 & =  & z_0 + \Delta s \sin\theta \nonumber \\
\rho_1 & = & \rho_0 + \Delta s \cos\theta
\eea

\noi
where $z_0 = c_2\xi_0\eta_0$, $\rho_0 = c_2 \sqrt{(\eta_0^2 - 1)(1 - \xi_0^2)}$ and $\tan\theta = z_0 (\eta_0^2 - 1)/(\eta_0^2 \rho_0)$.
The point ($\rho_1,z_1$) can be assumed to lie on another
hemiellipsoid $\eta_1 = \eta_0 + \Delta \eta$ and is defined alternately by the
co-ordinates ($\eta_1,\xi_1$) = ($\eta_0 + \Delta \eta, \xi_0 + \Delta \xi$) where $\Delta \eta$ and $\Delta \xi$ can be
computed by demanding that the point outside satisfies the ellipsoid/hyperboloid equation. Thus, $\Delta \eta$ is
determined using 

\be
\frac{z_1^2}{c_2^2 \eta_1^2} + \frac{\rho_1^2}{c_2^2(\eta_1^2 - 1)} = 1 \label{eq:defining1}
\ee

\noi
while $\Delta \xi$ can be evaluated either using

\be
\frac{z_1^2}{c_2^2 \xi_1^2} - \frac{\rho_1^2}{c_2^2(1 - \xi_1^2)} = 1 \label{eq:defining2}
\ee

\noi
or using $\xi_1 = (z_0 + \Delta s \sin\theta)/(c_2 \eta_1)$ and Eq.~\ref{eq:defining1}.
The solutions, to the accuracy required, can be expressed respectively as

\bea
\Delta \eta(\Delta s) & = & a_1 \Delta s + a_2 (\Delta s)^2 + a_3 (\Delta s)^3 + \mathcal{O}((\Delta s)^4) \label{eq:deta} \\
\Delta \xi(\Delta s) & = & b_1 \Delta s + b_2 (\Delta s)^2 + \mathcal{O}((\Delta s)^3) \label{eq:dxi}
\eea

\noi
where

\bea
a_1 & = & \frac{1}{h_\eta} \\
a_2 & = &  \frac{1}{2h_\xi^2} \frac{\eta_0}{\eta_0^2 - \xi_0^2} \\
a_3 & = & - \frac{1}{2 h_\eta h_\xi^2} \frac{\eta_0^2 + \xi_0^2}{(\eta_0^2 - \xi_0^2)^2}
\eea

\noi
while

\bea
b_1 & = & 0 \\
b_2 & = & \frac{1}{2h_\xi^2} \frac{\xi_0}{\eta_0^2 - \xi_0^2}
\eea

\noi
with

\bea
h_\eta & = & c_2 \sqrt{\frac{\eta_0^2 - \xi_0^2}{\eta_0^2 - 1}} \\
h_\xi & = & c_2 \sqrt{\frac{\eta_0^2 - \xi_0^2}{1 - \xi_0^2}}.
\eea

A Taylor expansion of the potential at the point ($\eta_0,\xi_0$) along the normal
can be expressed as

\be
\begin{aligned}
  V(\Delta s) = V_0 + V_\eta \Delta \eta  + V_\xi \Delta \xi + \frac{1}{2} V_{\eta\eta} (\Delta \eta)^2 + \\
  \frac{1}{2} V_{\xi\xi} (\Delta \xi)^2 + V_{\xi\eta} \Delta \eta \Delta \xi + \frac{1}{6} V_{\eta\eta\eta} (\Delta \eta)^3 + \ldots
  \end{aligned}
\ee

\noi
where $V_0 = V(\eta_0,\xi_0)$ and $\Delta \eta$ and $\Delta \xi$ are given by Eqns.~(\ref{eq:deta}) and (\ref{eq:dxi}) respectively.
Clearly, this expansion suffices to expand the potential upto $\mathcal{O}((\Delta s)^3)$ since $b_1$ is
zero. Also, since the $V_{\xi\xi}$ term contributes $\mathcal{O}((\Delta s)^4)$, it will be ignored
henceforth. The relevant partial derivatives can be evaluated as follows:

\bea
& V_\eta & =  -E_l h_\eta \\
& V_\xi & =  0 \\
& V_{\eta\eta}& =  E_l h_\eta~ \frac{ 2 \eta_0}{\eta_0^2 - 1}  \\
& V_{\xi\eta} & =  -E_l h_\eta~ \frac{1}{\xi_0} \\
& V_{\eta\eta\eta} &  =  -E_lh_\eta ~\frac{8\eta_0^2}{(\eta_0^2 - 1)^2}
\eea

\noi
Collecting together terms $\mathcal{O}((\Delta s)^k)$, the potential for the hemiellipsoid
is expressed as

\be
V(\Delta s) = V_0 + d_1 \Delta s + d_2 (\Delta s)^2 + d_3 (\Delta s)^3 + \mathcal{O}((\Delta s)^4)
\ee

\noi
where

\bea
d_1 &  = &  V_\eta a_1 \\
d_2 & = & V_\eta a_2 + \frac{1}{2} V_{\eta\eta} a_1^2 \\
d_3 & = & V_\eta a_3 + V_{\eta\eta} a_1 a_2 + V_{\xi\eta} a_1 b_2 + \frac{1}{6} V_{\eta\eta\eta} a_1^3
\eea

Consider now a sharp hemiellipsoid emitter $\eta = \eta_0$ for which $R_a/h << 1$ and
a point ($\eta_0,\xi_0$) on its surface at a height $z_0 = h - R_a/n$. At the apex, $n \rightarrow \infty$,
while the apex neighbourhood from where emission predominantly takes place corresponds generally to $n >> 10$.
The following approximations can then be made:

\bea
\eta_0 & = & \frac{h}{c_2} \simeq 1 + \frac{1}{2} \frac{R_a}{h} \\
\xi_0 & = & \frac{z_0}{h} \simeq 1 - \frac{1}{n} \frac{R_a}{h} \\
\eta_0^2 - 1& \simeq & \frac{R_a}{h} \\
1 - \xi_0^2 & \simeq & \frac{2}{n} \frac{R_a}{h} \\
\eta_0^2 - \xi_0^2 & \simeq & \frac{R_a}{h} (1 + \frac{2}{n}) \\
R_2 & = & R_a (1 + \frac{2}{n})^{1/2}
\eea

\noi
A comparison of the terms in $d_2$ at $E_l = 1$ yields

\bea
& V_\eta a_2 & =  \mathcal{O}(\frac{1}{n}\frac{1}{R_a}) \\
& V_{\eta\eta} a_1^2 & = \mathcal{O}(\frac{1}{R_a})
\eea

\noi
while the terms in $d_3$ at $E_l = 1$ are

\bea
& V_\eta a_3 & = \mathcal{O}(\frac{1}{n}\frac{1}{R_a^2}) \\
& V_{\xi\eta} a_1 b_2 & = \mathcal{O}(\frac{1}{n}\frac{1}{ h R_a}) \\
& V_{\eta\eta} a_1 a_2 & = \mathcal{O}(\frac{1}{n}\frac{1}{R_a^2})\\
& V_{\eta\eta\eta} a_1^3 & = \mathcal{O}(\frac{1}{R_a^2})
\eea

\noi
Thus, for a sharp emitter with $R_a/h << 1$, in the region close to the apex ($n >> 10$) from
where electron emission predominantly occurs at moderate fields,
$d_2 \simeq \frac{1}{2} V_{\eta\eta} a_1^2$ while
$d_3 \simeq \frac{1}{6} V_{\eta\eta\eta} a_1^3$. This leads to
Eq.~\ref{eq:ellip_final}.

In part therefore, the results obtained using Eq.~\ref{eq:ellip_final} are in good agreement because the
apex neighbourhood contributes substantially to the current. Our results show that at a local fields of 
$5$~V/nm, the region $n \geq 10$ contributes as much as 70\% to the total current. There is also a cancellation
of effects. The correction to the coefficient of the $(\Delta s/R_2)^2$ term leads to an increase in current while the correction
to the coefficent of the $(\Delta s/R_2)^3$ term leads to a decrease. Our studies for various field strengths and
apex radius of curvature, show that Eq.~\ref{eq:ellip_final} provides an optimum description of the external potential.

\vskip 0.05 in
$\;$\\
\section{References} 


\end{document}